\documentclass[10pt,aps,prb,twocolumn,amsmath,amssymb,superscriptaddress]{revtex4-1}
\usepackage{amsmath}
\usepackage{graphicx}
\usepackage{color}
\usepackage{multirow}
\usepackage[caption = false]{subfig}
\usepackage{array}
\usepackage{threeparttable}
\usepackage{mhchem}
\usepackage{mathtools}
\usepackage{siunitx}
\usepackage{centernot}
\usepackage{tikz}
\tikzset{font={\fontsize{11pt}{12}\selectfont}}
\usetikzlibrary{shapes,arrows}
\usepackage{lipsum}
\DeclareUnicodeCharacter{2032}{$^\prime$}
\makeatletter

\renewcommand*{\p@subsection}{}

\renewcommand*{\p@subsubsection}{}
\makeatother

\usepackage{hyperref}
\hypersetup{
	linkcolor=blue,          
	citecolor=blue,        
	urlcolor=blue,           
}

\newcommand{\ket}[1]{|#1\rangle}

\newcommand{\inner}[2]{\langle#1|#2\rangle}

\newcommand{\expecth}[3]{\langle#1|#2|#3\rangle}

\begin{document}
\title{Taming the sign problem in auxiliary field quantum Monte Carlo using accurate trial wave functions}

\author{Ankit Mahajan}
\email{ankitmahajan76@gmail.com}
\affiliation{Department of Chemistry, University of Colorado, Boulder, CO 80302, USA}

\author{Sandeep Sharma}
\affiliation{Department of Chemistry, University of Colorado, Boulder, CO 80302, USA}
\begin{abstract}
We explore different ways of incorporating accurate trial wave functions into free projection auxiliary field quantum Monte Carlo (fp-AFQMC). Trial states employed include coupled cluster singles and doubles, multi-Slater, and symmetry projected mean-field wave functions. We adapt a recently proposed fast multi-Slater local energy evaluation algorithm for fp-AFQMC, making the use of long expansions from selected configuration interaction methods feasible. We demonstrate how these trial wave functions serve to mitigate the sign problem and accelerate convergence in quantum chemical problems, allowing the application of fp-AFQMC to systems of substantial sizes. Our calculations on the widely studied model \ce{Cu2O2^2+} system show that many previously reported isomerization energies differ substantially from the near-exact fp-AFQMC value.
\end{abstract}
\maketitle

\section{Introduction}
Projector Monte Carlo (PMC) is among the most powerful and versatile methods for calculating properties of many-fermion systems.\cite{ceperley1977monte,nightingale1998quantum,foulkes2001quantum,becca2017quantum} In this method, one numerically integrates the imaginary-time ($\tau$) Schr\"odinger equation
\begin{equation}
	\frac{\partial|\psi(\tau)\rangle}{\partial\tau}=-\hat{H}|\psi(\tau)\rangle,
\end{equation}
whose solution can formally be written as $|\psi(\tau)\rangle=\exp(-\tau\hat{H})|\psi_{r}\rangle$, where \(\ket{\psi_r}\) is a trial state. Writing the trial state as a linear combination of the Hamiltonian eigenstates ($|\psi_{i}\rangle$), $|\psi_{r}\rangle=\sum_{i}c_{i}|\psi_{i}\rangle$ , we get
\begin{equation}
|\psi(\tau)\rangle=\sum_{i}c_{i}\exp(-E_{i}\tau)|\psi_{i}\rangle\label{eq:pmcsol},
\end{equation}
where \(E_i\) are energy eigenvalues. Consequently, as long as the initial state $|\psi_{r}\rangle$ does not have a vanishing overlap with the ground state ($c_{0}\neq0$), $|\psi(\tau)\rangle$ approaches the ground state exponentially fast in the long \(\tau\) limit. One can use another trial wave function $|\psi_l\rangle$ to measure various observables in the ground state. In particular, the ground state energy can be obtained using the expression
\begin{equation}
  E(\tau)=\frac{\langle\psi_{l}|He^{-\tau\hat{H}}|\psi_r\rangle}{\langle\psi_{l}|e^{-\tau\hat{H}}|\psi_r\rangle}\label{eq:estimator}
\end{equation}
which converges to the exact ground state energy in the long $\tau$ limit. Although PMC is exact in principle, in practice (in the absence of special symmetries in the system), any attempt at stochastic numerical integration is plagued with a severe numerical instability, termed the sign problem, stemming from the antisymmetry of fermionic wave functions.\cite{troyer2005computational} As a result, the signal-to-noise of the PMC simulation decreases exponentially, and after a relatively short imaginary time $\tau$, obtaining useful information from the simulation becomes intractable.

\begin{table}
\caption{Characteristics of various commonly used PMC approaches. The last column lists the bias that is almost always used to overcome the sign problem in the respective methods. \label{tab:pmc}}
\begin{tabular}{ccccccc}
\hline
Method &~~& Space &~~& Projector form &~~& Bias\\
\hline
DMC && Real && $\exp(-\Delta\tau H)$ && Fixed-node\\
GFMC && Real && $\left(1-\Delta\tau H\right)^{-1}$ && Fixed-node\\
GFMC && Orbital && $1-\Delta\tau H$ && Fixed-node\\
AFQMC && Orbital && $\exp(-\Delta\tau H)$ && Phaseless\\
FCIQMC && Orbital && $1-\Delta\tau H$ && Initiator\\
\hline
\end{tabular}
\end{table}

Various flavors of PMC differ in the space in which the simulation is performed, the method used to approximately apply the projector onto a state, and how the sign problem is controlled. Characteristics of four commonly used PMC methods, namely diffusion Monte Carlo (DMC),\cite{ceperley1980ground,foulkes2001quantum} Green's function Monte Carlo (GFMC),\cite{kalos1974helium,trivedi1990ground} auxiliary field QMC (AFQMC),\cite{sugiyama1986auxiliary,white1989numerical,sorella1989novel,hamann1990energy,zhang1997constrained,rom1997shifted,al2006auxiliary,motta2018ab} and full configuration interaction QMC (FCIQMC),\cite{BooThoAla,booth2013towards} are listed in table \ref{tab:pmc}. These approaches have been developed with different goals in mind based on their respective advantages and pitfalls. The sign problem is comparatively less severe in orbital space methods than real space methods due to the antisymmetry naturally enforced in the Hilbert space. The linear projectors used in orbital space GFMC and FCIQMC have the advantage that they do not have time step errors that arise in the exponential projectors. The cost of applying the linear projector is also much smaller comparatively. Alternatively, the exponential projector used in AFQMC allows noise-free application of the one-body part of the Hamiltonian, leading to a milder sign-problem in AFQMC compared to FCIQMC and GFMC. Regardless of the variation in the severity of the sign problem in different methods, it becomes exponentially worse with systems size in all cases, making calculations of large systems very difficult. One of the common ways to deal with it is to use a guiding wave function to enforce a constraint that stabilizes the simulation at the expense of a systematic bias. The phaseless constraint\cite{ZhaKra} often used in AFQMC (ph-AFQMC) and the fixed-node approximation used in DMC and GFMC lead to polynomially scaling methods, albeit with more or less controlled errors. A different approach, called the initiator approximation,\cite{CleBooAla} is used in FCIQMC, whereby the constraint is applied without an external guiding state using the initiator approximation. Another way of dealing with the sign problem is using transcorrelation, wherein trial wave function information gets folded into the Hamiltonian, making the projection more manageable.\cite{dobrautz2019compact} A recently proposed approach of combining aspects of variational and projection methods through time-step optimization also seems promising.\cite{sorella2021phase}

In the current work, we will take a different approach. Instead of introducing a bias to stabilize the simulation, we use high-quality trial states \(\ket{\psi_l}\) and $|\psi_{r}\rangle$ such that the simulation converges quickly enough before the sign problem becomes prohibitively expensive. We will use the exponential projector $\exp(-\Delta\tau H)$ and work in orbital space without enforcing a constraint. Thus this approach is almost identical to free projection AFQMC (fp-AFQMC). Looking at equation \ref{eq:estimator}, it is evident that if the trial states have substantial weight on the ground state and small weights on low lying excited states, the energy estimator \(E(\tau)\) will converge rapidly to the ground state energy, and one can obtain an accurate estimate of the ground state energy with a relatively short $\tau$ simulation. This approach is analogous to finite-temperature AFQMC simulations with \(\tau\) playing the role of inverse temperature. It is also closely related to the method presented in reference \onlinecite{hlubina1997ferromagnetism} by Sorella \textit{et al}., where a Gutzwiller trial wave function is used to obtain accurate energies in the Hubbard model by due to a zero-variance principle. While past efforts to incorporate high-quality wave functions into AFQMC, including those in ph-AFQMC,\cite{shi2014symmetry,wouters2014projector} have focused on model lattice systems, in this paper, we consider \textit{ab inito} systems and wave functions commonly used in quantum chemistry. We study the use of multi-Slater (obtained from selected configuration interaction methods),\cite{Huron1973,giner2013using,evangelista2014adaptive,Holmes2016b,tubman2016deterministic} coupled cluster singles and doubles (CCSD),\cite{vcivzek1966correlation,bartlett2007coupled} and symmetry projected mean-field states\cite{scuseria2011projected,jimenez2012projected} as trial wave functions in fp-AFQMC. In particular, we find that multi-Slater states, in combination with CCSD, give remarkably accurate results for several systems, and one can perform near-exact simulations on systems as large as 52 electrons in 118 orbitals.

The rest of this paper is organized as follows: In section \ref{sec:theory}, we begin with the details of the procedure used for sampling the short-time propagator. Then we present efficient ways of using various trial states in the estimator. In section \ref{sec:results}, we analyze the efficacy of these states in fp-AFQMC with some illustrative examples. We also gauge the accuracy of focal point corrections to stretch the scope of the method. We present results for larger systems, namely benzene, cyclobutadiene, and \ce{Cu2O2^2+}. Finally, we conclude with a discussion of the prospects for our approach.

\section{Theory}
\label{sec:theory}
\subsection{Propagator sampling}
\label{sec:propagator_sampling}
Here, we review how the propagator is sampled in AFQMC following the detailed exposition by \citet{motta2018ab}. Consider the \textit{ab initio} electronic Hamiltonian given by
\begin{equation}
   \hat{H} = \sum_{ij}h^{j}_{i}a_i^{\dagger}a_j + \frac{1}{2}\sum_{ikjl}v^{kl}_{ij}a_i^{\dagger}a_ka_j^{\dagger}a_l = \hat{K} + \hat{V},\label{eq:ham}
\end{equation}
where \(i, j, k, l\) are spin orbital indices, and \(h^{j}_{i}\) and \(v^{kl}_{ij}\) are one and two electron integrals, respectively. We denote the number of orbitals by \(M\) and the number of electrons by \(N\). To sample the imaginary time propagator, one first divides the propagation into small intervals and makes a Trotter approximation for each interval, given as
 \begin{equation}
   \begin{split}
       e^{-\tau \hat{H}} &= \left(e^{-\Delta\tau \hat{H}}\right)^N,\\
			 e^{-\Delta\tau \hat{H}} &= e^{-\frac{\Delta\tau \hat{K}}{2}}e^{-\Delta\tau \hat{V}}e^{-\frac{\Delta\tau \hat{K}}{2}} + O(\Delta\tau^3).
   \end{split}
\end{equation}

To sample the exponential of the two-body part, we express it as a sum of squares using a modified Cholesky decomposition:\cite{beebe1977simplifications,koch2003reduced}
\begin{equation}
   \hat{V} = \frac{1}{2}\sum_{\gamma} \left(\sum_{jl}\left[L^{\gamma}\right]^{l}_{j}a_j^{\dagger}a_l\right)^2 = -\frac{1}{2}\sum_{\gamma} \hat{v}_{\gamma}^2,
\end{equation}
where \(\hat{v}_{\gamma} = i\sum_{jl} \left[L^{\gamma}\right]^{l}_{j}a_j^{\dagger}a_l\) are one-body operators. We will denote the number of Cholesky matrices, \(L^{\gamma}\), by \(X\). Empirically, \(X\) is known to scale linearly with the number of orbitals with \(X \sim 5\)-\(10 M\). Using the Hubbard-Stratanovic transform, we get
\begin{equation}
   e^{-\Delta\tau\hat{V}} = \prod_{\gamma}\int \frac{dx_{\gamma}}{\sqrt{2\pi}}e^{-\frac{x_{\gamma}^2}{2}}e^{\sqrt{\Delta\tau}x_{\gamma}\hat{v}_{\gamma}} + O(\Delta\tau^2),
\end{equation}
where \(x_{\gamma}\) are the auxiliary fields. The quadratic Trotter error arises due to the fact that \(\hat{v}_{\gamma}\) do not commute with each other. The propagator on the short interval can be expressed in a compact form as
\begin{equation}
   e^{-\Delta\tau \hat{H}} = \int d \mathbf{x}\ p(\mathbf{x})\hat{B}(\mathbf{x}) + O(\Delta\tau^2),
\end{equation}
where \(\mathbf{x}\) is the vector of auxiliary fields, \(p(\mathbf{x})\) is the standard Gaussian distribution and \(\hat{B}(\mathbf{x})\) is a one-body operator given by
\begin{equation}
   \hat{B}(\mathbf{x}) = e^{-\frac{\Delta\tau K}{2}}e^{\sqrt{\Delta\tau}\sum_{\gamma}x_{\gamma}\hat{v}_{\gamma}}e^{-\frac{\Delta\tau K}{2}}.\label{eq:prop}
\end{equation}
The full propagator can be written as
\begin{equation}
   e^{-\tau \hat{H}} = \mathcal{T}\prod_{i}\int d \mathbf{x}_i\ p(\mathbf{x}_i)\hat{B}(\mathbf{x}_i) = \int dX\  p(X)\hat{\mathcal{B}}(X),
\end{equation}
where \(\mathbf{x}_i\) are auxiliary fields at the \(i\)th time slice, \(\mathcal{T}\) denotes time ordering, \(X = \left\{\mathbf{x}_i\right\}\) is the set of auxiliary fields at all time slices, and \(\hat{\mathcal{B}}(X) = \mathcal{T}\prod_i\hat{B}(\mathbf{x}_i)\).

Substituting the propagator integral into the energy expression in equation \ref{eq:estimator}, we get
\begin{equation}
   E(\tau) = \frac{\int dX\ p(X)\expecth{\psi_l}{\hat{H}\hat{\mathcal{B}}(X)}{\psi_r}}{\int dX\ p(X)\expecth{\psi_l}{\hat{\mathcal{B}}(X)}{\psi_r}}.
\end{equation}
According to the Thouless theorem, applying the exponential of a one-body operator onto a Slater determinant results in another Slater determinant with rotated orbitals. This allows one to operate the sampled propagator onto a determinant as
\begin{equation}
   \hat{\mathcal{B}}(X)\ket{\phi} = \ket{\phi(X)},
\end{equation}
where \(\ket{\phi}\) and \(\ket{\phi(X)}\) are Slater determinants. If the trial state is a single determinant, the projector can thus be applied directly to it. But even when the trial state \(\ket{\psi_r}\) is not a single determinant (e.g. CCSD or Jastrow-Slater wave function), it can be written as
\begin{equation}
   \ket{\psi_r} = \int d\phi\ c(\phi)\ket{\phi},\label{eq:samplePsi}
\end{equation}
where the integral is over an overcomplete set of Slater determinants \(\ket{\phi}\), and consequently, the expansion coefficients \(c(\phi)\) are not unique. Thus we have
\begin{equation}
   E(\tau) = \frac{\int dXd\phi\ p(X)c(\phi)\expecth{\psi_l}{\hat{H}\hat{\mathcal{B}}(X)}{\phi}}{\int dXd\phi\ p(X)c(\phi)\expecth{\psi_l}{\hat{\mathcal{B}}(X)}{\phi}}.
\end{equation}
Now we can sample the multidimensional integrals in the numerator and the denominator using Monte Carlo. The process for directly sampling determinants \(\ket{\phi}\) from the trial state \(\ket{\psi_r}\) will be described in section \ref{sec:trial_wave_functions} for a few different types of trials. We sample \(X\) directly from the standard Gaussian distribution \(p(X)\). So we get the estimator
\begin{equation}
   E(\tau) \approx \frac{\sum_{i}\expecth{\psi_l}{\hat{H}\hat{\mathcal{B}}(X_i)}{\phi_i}}{\sum_{i}\expecth{\psi_l}{\hat{\mathcal{B}}(X_i)}{\phi_i}} = \frac{\sum_{i}\expecth{\psi_l}{\hat{H}}{\phi_i(X_i)}}{\sum_{i}\inner{\psi_l}{\phi_i(X_i)}},\label{eq:zv}
\end{equation}
where \(X_i\) and \(\ket{\phi_i}\) are auxiliary field and trial wave function samples, respectively. We note that this estimator of the ratio of two random variables has a bias that goes down as \(1/N_{\text{samples}}\), while the statistical noise goes down more slowly as \(1/\sqrt{N_{\text{samples}}}\). Therefore we ignore the more rapidly decaying bias, which we have confirmed to be unimportant in our numerical calculations. In the limit \(\ket{\psi_l}\) is the exact ground state, the estimator in equation \ref{eq:zv} becomes noiseless since \(\expecth{\psi_0}{\hat{H}}{\phi_i(X_i)}=E_0\inner{\psi_0}{\phi_i(X_i)}\). This zero-variance property ensures more accurate \(\ket{\psi_l}\) lead to lower statistical noise in the estimator. Subtracting the mean-field background from the Hamiltonian is known to substantially reduce the statistical fluctuations,\cite{rom1997shifted} and we employ this strategy in our calculations. We also periodically orthogonalize the orbitals in the walker, \(\ket{\phi_i(X_i)}\), during the propagation at fixed intervals for numerical stability.

The most expensive operation in propagation is the formation of the one-body operator matrix (\(\sum_{\gamma}x_{\gamma}\hat{v}_{\gamma}\) in equation \ref{eq:prop}), which has a computational cost scaling of \(O(XM^2)\). The exponentials of these matrices are not explicitly calculated, instead only their action on the walker matrices, \(\ket{\phi_i(X_i)}\) with dimension \(M\times N\), is calculated using a truncated Taylor series, which has a cost scaling of \(O(NM^2)\). We retain the first ten terms in the Taylor series of the exponential.

\subsection{Trial wave functions}
\label{sec:trial_wave_functions}
In this section, we will outline methods for using various accurate trial states as \(\ket{\psi_l}\) and \(\ket{\psi_r}\) in equation \ref{eq:estimator}. For using a wave function as \(\ket{\psi_l}\), one needs an efficient way to calculate its local energy given by
\begin{equation}
   E_L(\phi) = \frac{\expecth{\psi_l}{\hat{H}}{\phi}}{\inner{\psi_l}{\phi}},
\end{equation}
where \(\ket{\phi}\) is a Slater determinant. In terms of local energy, equation \ref{eq:zv} is written as
\begin{equation}
   E(\tau) \approx \frac{\sum_i\inner{\psi_l}{\phi_i(X_i)}E_L(\phi_i(X_i))}{\sum_i\inner{\psi_l}{\phi_i(X_i)}}.
\end{equation}
For accurate trial states and larger systems, local energy evaluation can become more expensive than propagation. Many ways of reducing this cost in AFQMC have been proposed for single determinant and multi-Slater trials.\cite{malone2018overcoming,shee2018phaseless,motta2019efficient,lee2020stochastic,shi2021some} Below we will describe economic algorithms for evaluating local energies for multi-Slater and symmetry projected mean-fields trial states which can be used as \(\ket{\psi_l}\). To use a wave function as the trial state \(\ket{\psi_r}\) in the sampling method outlined in the last section, one needs to be able to sample a determinant from it (equation~\ref{eq:samplePsi}). We will present a way to do this for CCSD wave functions.

\subsubsection{Multi-Slater}
\label{sec:multi_slater}
A multi-Slater wave function is obtained by particle-hole excitations from a reference configuration. We can write it as
\begin{equation}
   \ket{\psi} = \sum_{n}^{N_c}c_n\prod_{\mu_i}^{k_n}a^{\dagger}_{t_{\mu_n}}a_{p_{\mu_n}}\ket{\psi_0},
\end{equation}
where \(\ket{\psi_0}\) is the reference configuration, \(c_n\) are real expansion coefficients, \(N_c\) is the number of configurations in the expansion, and \(k_n\) are the excitation ranks. Note the that the product of particle-hole excitations need not be ordered since the excitations commute with each other.

Multi-Slater wave functions can be employed as \(\ket{\psi_l}\) trial states in fp-AFQMC. We recently presented an efficient local energy evaluation algorithm for multi-Slater wave functions in variational Monte Carlo\cite{mahajan2020efficient} that has a favorable scaling with the number of configurations, analogous to developments in real space QMC algorithms.\cite{Assaraf2017} Here this algorithm is adapted for AFQMC with some important changes. For convenience, we write the local energy as
\begin{equation}
   E_L[\phi] = \frac{\expecth{\psi}{\hat{H}}{\phi}}{\inner{\psi_0}{\phi}} \Big/  \frac{\inner{\psi}{\phi}}{\inner{\psi_0}{\phi}}.\label{eq:multi_eloc}
\end{equation}
The overlap ratio in the denominator is given by
\begin{equation}
   \frac{\inner{\psi}{\phi}}{\inner{\psi_0}{\phi}} = \sum_n c_n \frac{\expecth{\psi_0}{\prod_{\mu_n}a^{\dagger}_{p_{\mu_n}}a_{t_{\mu_n}}}{\phi}}{\inner{\psi_0}{\phi}}.\label{eq:overlap}
\end{equation}
Since \(\ket{\psi_0}\) and \(\ket{\phi}\) are determinants, we can use the generalized Wick's theorem to compute the terms in this sum. The Green's function is given as\cite{motta2018ab}
\begin{equation}
   G^i_j = \frac{\expecth{\psi_0}{a^{\dagger}_ia_j}{\phi}}{\inner{\psi_0}{\phi}} = \left[\phi(\psi_0^{\dagger}\phi)^{-1}\psi_0^{\dagger}\right]^j_i,
\end{equation}
where \(\phi\) and \(\psi_0\) are the orbital coefficient matrices of the corresponding determinants. Using the generalized Wick's theorem, we get
\begin{equation}
   \frac{\expecth{\psi_0}{\prod_{\mu}a^{\dagger}_{p_{\mu}}a_{t_{\mu}}}{\phi}}{\inner{\psi_0}{\phi}} = \det \left(G^{\left\{p_{\mu}\right\}}_{\left\{t_{\mu}\right\}}\right),
\end{equation}
where the sets of indices \(\left\{p_{\mu}\right\}\) and \(\left\{t_{\mu}\right\}\) denote the \(k\times k\) slice of the \(G\) matrix, \(k\) being the rank of the excitation. Therefore once the Green's function is calculated at cost \(O(N^2M)\) the overlap ratio in equation \ref{eq:overlap} can be calculated at cost \(O(N_c)\). We will ignore the scaling factors with respect to the excitation rank \(k\) as it is usually small, typically (\(k \leq 6\)).

To calculate the numerator in the local energy expression of equation \ref{eq:multi_eloc}, we first rewrite the Hamiltonian as
\begin{equation}
   \hat{H} = \sum_{ij}\left[h'\right]_{j}^{i}a_i^{\dagger}a_j + \frac{1}{2}\sum_{ikjl}\sum_{\gamma}\left[L^{\gamma}\right]^{k}_{i}\left[L^{\gamma}\right]^{l}_{j}a_i^{\dagger}a_j^{\dagger}a_ka_l,
\end{equation}
where we have normal-ordered the two-body term in equation \ref{eq:ham} and absorbed the resulting one-body terms into \(h'\). Note that we have used the same orbitals as those used in the multi-Slater trial wave function to express the Hamiltonian. The one body part of the local energy numerator is given by
 \begin{equation}
   \begin{split}
     \frac{\expecth{\psi}{\hat{H}_1}{\phi}}{\inner{\psi_0}{\phi}} =&\\
		 \sum_n& c_n \left(\sum_{ij}\left[h'\right]^{j}_{i}\frac{\expecth{\psi_0}{\left(\prod_{\mu_n}a^{\dagger}_{p_{\mu_n}}a_{t_{\mu_n}}\right)a_i^{\dagger}a_j}{\phi}}{\inner{\psi_0}{\phi}}\right)\label{eq:onebody}
   \end{split}
\end{equation}
Using the generalized Wick's theorem, we get
\begin{equation}
   \frac{\expecth{\psi_0}{\left(\prod_{\mu}a^{\dagger}_{p_{\mu}}a_{t_{\mu}}\right)a_i^{\dagger}a_j}{\phi}}{\inner{\psi_0}{\phi}} = \det
	 \begin{pmatrix}
		 G^i_j & \mathcal{G}^i_{\left\{t_{\mu}\right\}}\\[1em]
		 G^{\left\{p_{\mu}\right\}}_j & G^{\left\{p_{\mu}\right\}}_{\left\{t_{\mu}\right\}}\\
	 \end{pmatrix},\label{eq:green1}
\end{equation}
where the matrix on the RHS is written as a block matrix with the blocks given by the specified slices. We have also defined a modified Green's function \(\mathcal{G}\) as
\begin{equation}
   \mathcal{G}^i_j = G^i_j - \delta^i_j,
\end{equation}
where \(\delta\) is the Kronecker delta. A naive evaluation of the sums  in equation \ref{eq:onebody} entails contracting over the one-body Hamiltonian separately for each configuration in the multi-Slater expansion. This explicit evaluation of the double sum can be avoided by using the following trick. Substituting the matrix element in equation \ref{eq:green1} into equation \ref{eq:onebody}, and dropping the configuration index \(n\) for convenience, we get
\begin{equation}
	\begin{split}
		\sum_{ij}\left[h'\right]^{j}_{i} & \det
		\begin{pmatrix}
			G^i_j & \mathcal{G}^i_{\left\{t_{\mu}\right\}}\\[1em]
			G^{\left\{p_{\mu}\right\}}_j & G^{\left\{p_{\mu}\right\}}_{\left\{t_{\mu}\right\}}\\
		\end{pmatrix} = \\
		&\left(\sum_{ij}\left[h'\right]^{j}_{i}G^i_j\right)\det \left(G^{\left\{p_{\mu}\right\}}_{\left\{t_{\mu}\right\}}\right) \\
		& + \sum_{\nu=1}^k(-1)^{\nu}\det
		\begin{pmatrix}
			\sum_{ij}\left[h'\right]^j_iG^{p_\nu}_j\mathcal{G}^i_{\left\{t_{\mu}\right\}}\\[1em]
			G^{\left\{p_{\mu}\right\}\symbol{92}p_{\nu}}_{\left\{t_{\mu}\right\}}\\
		\end{pmatrix},\label{eq:laplace1}
	\end{split}
\end{equation}
where we have Laplace expanded the determinant on the LHS along the first column. \(\left\{p_{\mu}\right\}\symbol{92}p_{\nu}\) denotes the set of indices \(\left\{p_{\mu}\right\}\) excluding \(p_{\nu}\). This equation suggests a way of separating the sums over Hamiltonian and configuration indices by precomputing the intermediates given by
 \begin{equation}
   \begin{split}
		 E^1_0 &= \sum_{ij}\left[h'\right]^{j}_{i}G^i_j,\\
		 S^p_t &= \sum_{ij} \left[h'\right]^{j}_{i}G^p_j\mathcal{G}^i_t.\\
   \end{split}
\end{equation}
Using these intermediates in equation \ref{eq:laplace1}, we get
 \begin{equation}
   \begin{split}
		 \sum_{ij}\left[h'\right]^{j}_{i}  \det
		 \begin{pmatrix}
			 G^i_j & \mathcal{G}^i_{\left\{t_{\mu}\right\}}\\[1em]
			 G^{\left\{p_{\mu}\right\}}_j & G^{\left\{p_{\mu}\right\}}_{\left\{t_{\mu}\right\}}\\
		 \end{pmatrix} &=
		  E_0^1\det \left(G^{\left\{p_{\mu}\right\}}_{\left\{t_{\mu}\right\}}\right) \\
		 + &\sum_{\nu=1}^k(-1)^{\nu}\det
		 \begin{pmatrix}
			 S^{p_\nu}_{\left\{t_{\mu}\right\}}\\[1em]
			 G^{\left\{p_{\mu}\right\}\symbol{92}p_{\nu}}_{\left\{t_{\mu}\right\}}\\
		 \end{pmatrix},
   \end{split}
\end{equation}
where the RHS now does not involve any contractions over the one-body Hamiltonian indices.

%

The two-body contribution to the local energy numerator is given by
 \begin{equation}
    \frac{\expecth{\psi}{\hat{H}_2}{\phi}}{\inner{\psi_0}{\phi}} = \sum_n c_n E_n
	\end{equation}
	where
 \begin{equation}
   \begin{split}
    E_n&=\\ &\sum_{\gamma}\sum_{ikjl}\left[L^{\gamma}\right]^{k}_{i}\left[L^{\gamma}\right]^{l}_{j}\frac{\expecth{\psi_0}{\left(\prod_{\mu_n}a^{\dagger}_{p_{\mu_n}}a_{t_{\mu_n}}\right)a_i^{\dagger}a_j^{\dagger}a_ka_l}{\phi}}{\inner{\psi_0}{\phi}}.\label{eq:twobody}
   \end{split}
\end{equation}
Again using Wick's theorem, we find
 \begin{equation}
   \begin{split}
     \frac{\expecth{\psi_0}{\left(\prod_{\mu_n}a^{\dagger}_{p_{\mu_n}}a_{t_{\mu_n}}\right)a_i^{\dagger}a_j^{\dagger}a_ka_l}{\phi}}{\inner{\psi_0}{\phi}}& =\\
		 \det&
		 \begin{pmatrix}
			 G^{\left\{i,j\right\}}_{\left\{k,l\right\}} & \mathcal{G}^{\left\{i,j\right\}}_{\left\{t_{\mu}\right\}}\\[1em]
			 G^{\left\{p_{\mu}\right\}}_{\left\{k,l\right\}} & G^{\left\{p_{\mu}\right\}}_{\left\{t_{\mu}\right\}}\\
		 \end{pmatrix}.\label{eq:green2}
   \end{split}
\end{equation}
By following a procedure analogous to the one-body case, it is possible avoid the expensive explicit evaluation of the sums in equation \ref{eq:twobody}, by precomputing the following intermediates:
 \begin{equation}
   \begin{split}
		 E^2_0 & = \sum_{\gamma}\sum_{ikjl}\left[L^{\gamma}\right]^k_i \left[L^{\gamma}\right]^l_j\det \left(G^{\left\{i,j\right\}}_{\left\{k,l\right\}}\right)\\
		 \left[D_1\right]^p_t &= \sum_{\gamma}\sum_{ikjl}\left[L^{\gamma}\right]^{k}_{i}\left[L^{\gamma}\right]^{l}_{j}\det \left(G^{\left\{i,p\right\}}_{\left\{k,l\right\}}\right)\mathcal{G}^j_t,\\
		 \left[D_2^{\gamma}\right]^p_t &= \sum_{ik} \left[L^{\gamma}\right]^{k}_{i}G^p_k \mathcal{G}^i_t.\label{eq:intermediate}
   \end{split}
\end{equation}
Using these intermediates along with equations \ref{eq:twobody} and \ref{eq:green2}, dropping the configuration index \(n\) for convenience, we get
 \begin{equation}
   \begin{split}
    &\sum_{\gamma ikjl}\left[L^{\gamma}\right]^{k}_{i}\left[L^{\gamma}\right]^{l}_{j}\det
		\begin{pmatrix}
			G^{\left\{i,j\right\}}_{\left\{k,l\right\}} & \mathcal{G}^{\left\{i,j\right\}}_{\left\{t_{\mu}\right\}}\\[1em]
			G^{\left\{p_{\mu}\right\}}_{\left\{k,l\right\}} & G^{\left\{p_{\mu}\right\}}_{\left\{t_{\mu}\right\}}\\
		\end{pmatrix}=\\
				&\qquad\qquad E_0^2 \det \left(G^{\left\{p_{\mu}\right\}}_{\left\{t_{\mu}\right\}}\right)
				+\sum_{\nu=1}^k(-1)^{\nu}\det
	 		 \begin{pmatrix}
	 			 \left[D_1\right]^{p_\nu}_{\left\{t_{\mu}\right\}}\\[1em]
	 			 G^{\left\{p_{\mu}\right\}\symbol{92}p_{\nu}}_{\left\{t_{\mu}\right\}}\\
	 		 \end{pmatrix}\\
			 &+\sum_{\gamma}\sideset{}{'}\sum_{\nu_1,\nu_2,\lambda_1,\lambda_2}^k(-1)^{\nu_1+\nu_2+\lambda_1+\lambda_2}\\
			 &\qquad\qquad\times\det \left(\left[D_2^{\gamma}\right]^{\left\{p_{\nu_1},p_{\nu_2}\right\}}_{\left\{t_{\lambda_1,\lambda_2}\right\}}\right)\det \left(G^{\left\{p_{\mu}\right\}\symbol{92}\left\{p_{\nu_1},p_{\nu_2}\right\}}_{\left\{t_\mu\right\}\symbol{92}\left\{t_{\lambda_1},t_{\lambda_2}\right\}}\right),\label{eq:laplace}
   \end{split}
\end{equation}
where we Laplace expanded the determinant on the LHS along the first two columns. The primed sum indicates the conditions \(\nu_1\neq\nu_2\) and \(\lambda_1\neq\lambda_2\), and all these indices can take values from 1 to \(k\). Note that the RHS only involves a sum over the Hamiltonian index \(\gamma\), while all other Hamiltonian indices are contracted in the separate calculation of the intermediates. Thus we have partially separated the sums over configuration and Hamiltonian indices.

Now we consider the cost scaling of this algorithm. We will only look at the more expensive two-body contribution. The cost of calculating \(G\) scales as \(O(N^2M)\). The intermediate \(D_1\) in equation \ref{eq:intermediate} can be calculated at cost \(O(XNM^2)\). The cost of calculating each \(D_2^{\gamma}\) is \(O(NM^2)\), thus the cost of calculating it for all \(\gamma\) is also \(O(XNM^2)\). Given these intermediates, the contribution of each determinant in the expansion to the local energy, as given by equation \ref{eq:laplace}, has a cost scaling of \(O(X)\) arising from the last term in the equation. Therefore the total cost scaling is given by \(O(XNM^2 + XN_c)\). We note that this scaling is different from that of the algorithm we reported previously in reference \onlinecite{mahajan2020efficient}. It is possible to use our previous algorithm in fp-AFQMC, which amounts to combining the \(D_2^{\gamma}\) intermediates to build a four index tensor intermediate, resulting in a total cost scaling of \(O(XN^2M^2 + N_c)\). While this algorithm has a better scaling with \(N_c\), for the multi-Slater expansions employed in this study with \(N_c\sim 10^4\), we found it to be slower than the algorithm detailed above in most cases because of the large cost of building the four index intermediate. Finally, if the multi-Slater expansion is restricted to an active orbital space of size \(A\), the algorithm cost scales as \(O(XNAM + XN_c)\).

\subsubsection{CCSD}
\label{sec:ccsd}
The coupled cluster wave function ansatz is given by
 \begin{equation}
   \begin{split}
       \ket{\psi} &= \exp \left(\hat{T}_1+\hat{T}_2\right)\ket{\psi_0}\\
			  &= \exp \left(\sum_{\mu}t_{\mu}\hat{E}_{\mu} + \sum_{\mu,\nu}t_{\mu\nu}\hat{E}_{\mu}\hat{E}_{\nu}\right)\ket{\psi_0},
   \end{split}
\end{equation}
where \(\ket{\psi_0}\) is a reference Slater determinant, \(\hat{E}_{\mu}\) are particle-hole excitation operators given by
\begin{equation}
   \hat{E}_{\mu} = a_{t}^{\dagger}a_p.
\end{equation}
\(\mu = \{t, p\}\) is a composite index for a pair of empty and occupied orbitals in the reference determinant. \(t_{\mu}\) and \(t_{\mu\nu}\) are singles and doubles amplitudes, respectively. To use a CCSD wave function as \(\ket{\psi_r}\), we sample determinants from it by using a Hubbard-Stratanovic transform in a manner analogous to the imaginary time propagator. Since the particle-hole excitations operators commute with each other, we have
\begin{equation}
   \exp \left(\hat{T}_1+\hat{T}_2\right)\ket{\psi_0} = \exp \left(\hat{T}_2\right)\ket{\tilde{\psi}_0},
\end{equation}
where, due to Thouless' theorem, \(\ket{\tilde{\psi}_0} = \exp \left(\hat{T}_1\right)\ket{\psi_0}\) is a Slater determinant. Note that unlike the imaginary time propagator, here we can split the exponential exactly without incurring a Trotter error. Now we write \(\hat{T}_2\) as a sum of squares by diagonalizing the doubles amplitudes:
\begin{equation}
   \hat{T_2} = \sum_{\mu,\nu}t_{\mu\nu} \hat{E}_{\mu}\hat{E}_{\nu} = \sum_{i}\hat{\mathcal{E}}_i^2,
\end{equation}
where the one-body operators \(\hat{\mathcal{E}}_i\) are defined as
\begin{equation}
   \hat{\mathcal{E}}_i = \sqrt{\lambda_i}\sum_{\mu}V^{\mu}_i\hat{E}_{\mu}.
\end{equation}
Here \(\lambda_i\) and \(V\) are the eigenvalues and eigenvectors of the doubles amplitudes \(t_{\mu\nu}\), respectively. Using the Hubbard-Stratanovic transform, we get
\begin{equation}
   \exp \left(\sum_i \hat{\mathcal{E}}_i^2\right) = \prod_{i}\int \frac{dx_i}{\sqrt{2\pi}} e^{-\frac{x_i^2}{2}}e^{\sqrt{2}x_i\hat{\mathcal{E}}_i}.
\end{equation}
Having written the exponential of a two-body operator as an integral over exponentials of one-body operators, we can sample its action upon \(\ket{\tilde{\psi}_0}\) using Monte Carlo. We again employ direct sampling of the multi-dimensional Gaussian to obtain determinant samples (see section \ref{sec:propagator_sampling}). In this sense, the CCSD operator can be thought of as performing a long-time propagation of the reference determinant in one step without a Trotter error. Computational cost considerations for sampling the CCSD wave function this way are similar to those for sampling the propagator. We note that this way of sampling CCSD wave functions is similar to the treatment of Jastrow factors in reference \onlinecite{chang2016auxiliary}.

\subsubsection{Symmetry projected mean-field states}
Symmetry projected mean-field states are given by
\begin{equation}
   \ket{\psi} = \prod_{i}\hat{P}_i\ket{\psi_0},
\end{equation}
where \(\hat{P}_i\) denotes the projector onto an eigenstate of the operator for symmetry \(i\), and \(\ket{\psi_0}\) is a broken symmetry mean-field state. When such states are variationally optimized in the presence of the projectors they yield more accurate approximations of the desired energy eigenstate at a mean field level compared to bare symmetry broken states without projectors. The state \(\ket{\psi_0}\) can be allowed to break different symmetries of the molecular Hamiltonian like spin, complex conjugation, point group, and number symmetry simultaneously. Here we consider the breaking and restoration of spin (S) and complex conjugation (K) symmetries resulting in a KSGHF state. Details of optimizing such states can be found in reference \onlinecite{scuseria2011projected}.

To use KSGHF as the \(\ket{\psi_l}\) trial state in fp-AFQMC calculations, we consider the local energy given by
\begin{equation}
   E_L\left[\phi\right] = \frac{\expecth{\psi_0}{\hat{P}_K\hat{P}_S\hat{H}}{\phi}}{\expecth{\psi_0}{\hat{P}_K\hat{P}_S}{\phi}}.
\end{equation}
Note that \(\hat{P_S}\) commutes with the Hamiltonian. If determinants sampled from \(\ket{\psi_r}\) are closed shell (RHF-like) or high-spin open shell (ROHF-like), the propagator sampling described in section \ref{sec:propagator_sampling} preserves their good spin quantum numbers. So, in these cases, we have
\begin{equation}
   \hat{P}_S\ket{\phi} = \ket{\phi}.
\end{equation}
This allows us to ignore the spin projector completely because of its trivial action on the walker. If a sampling of \(\ket{\psi_r}\) were used that produced UHF-like determinants, one would have to perform the spin projection explicitly, but we will not consider this case here. The complex conjugation symmetry projection is performed as
\begin{equation}
   E_L\left[\phi\right] = \frac{\expecth{\psi_0}{\hat{H}}{\phi}+\expecth{\psi_0^*}{\hat{H}}{\phi}}{\inner{\psi_0}{\phi}+\inner{\psi_0^*}{\phi}},
\end{equation}
where \(\ket{\psi_0^*}\) is obtained by complex conjugating the orbitals of \(\ket{\psi_0}\). We use this same projector when optimizing the KSGHF state. Evaluation of local energy now proceeds similarly to the RHF case with a cost scaling of \(O(XN^2M)\).

\section{Results}\label{sec:results}
In this section, we present and analyze the results of fp-AFQMC calculations on various systems. We used PySCF\cite{sun2018pyscf} to obtain molecular electronic integrals and optimized CCSD wave functions. The modified Cholesky decompositions were performed using a script in the AFQMC code PAUXY.\cite{pauxy} HCI calculations were performed with the code Dice.\cite{ShaHolUmr,smith2017cheap} All fp-AFQMC calculations were performed with a code that we have made publicly available on GitHub.\onlinecite{dqmccode}. Input and output files from all large calculations can also be found in a public repository.\onlinecite{calcFiles}. Additional details of the calculations have been provided in the supporting information (SI).

Before presenting the results, we make a note of some practical details related to fp-AFQMC calculations. The amount of time required to achieve convergence depends on the gap to low-lying excited states and the convergence threshold. We estimated convergence by performing energy calculations at multiple points during the projection, and energies within error bars of each other at 2-3 consecutive substantially separated times were taken to be converged. All fp-AFQMC calculations have a Trotter error that can be made arbitrarily small by reducing the time step. There are various ways of handling this error common to almost all projection QMC methods. We assessed the amount of Trotter error by performing pilot calculations with different time steps for a fraction of the propagation time required for convergence. Based on these small calculations, conservative time steps were chosen. We employed time steps ranging from 0.01-0.1 H\(^{-1}\) in this study for different problems.

\subsection{Illustrative examples}\label{sec:illustrative_examples}
\subsubsection{Energy convergence for different trial states}\label{sec:energy_convergence}
\begin{figure}[htp]
\centering
\includegraphics[width=0.45\textwidth]{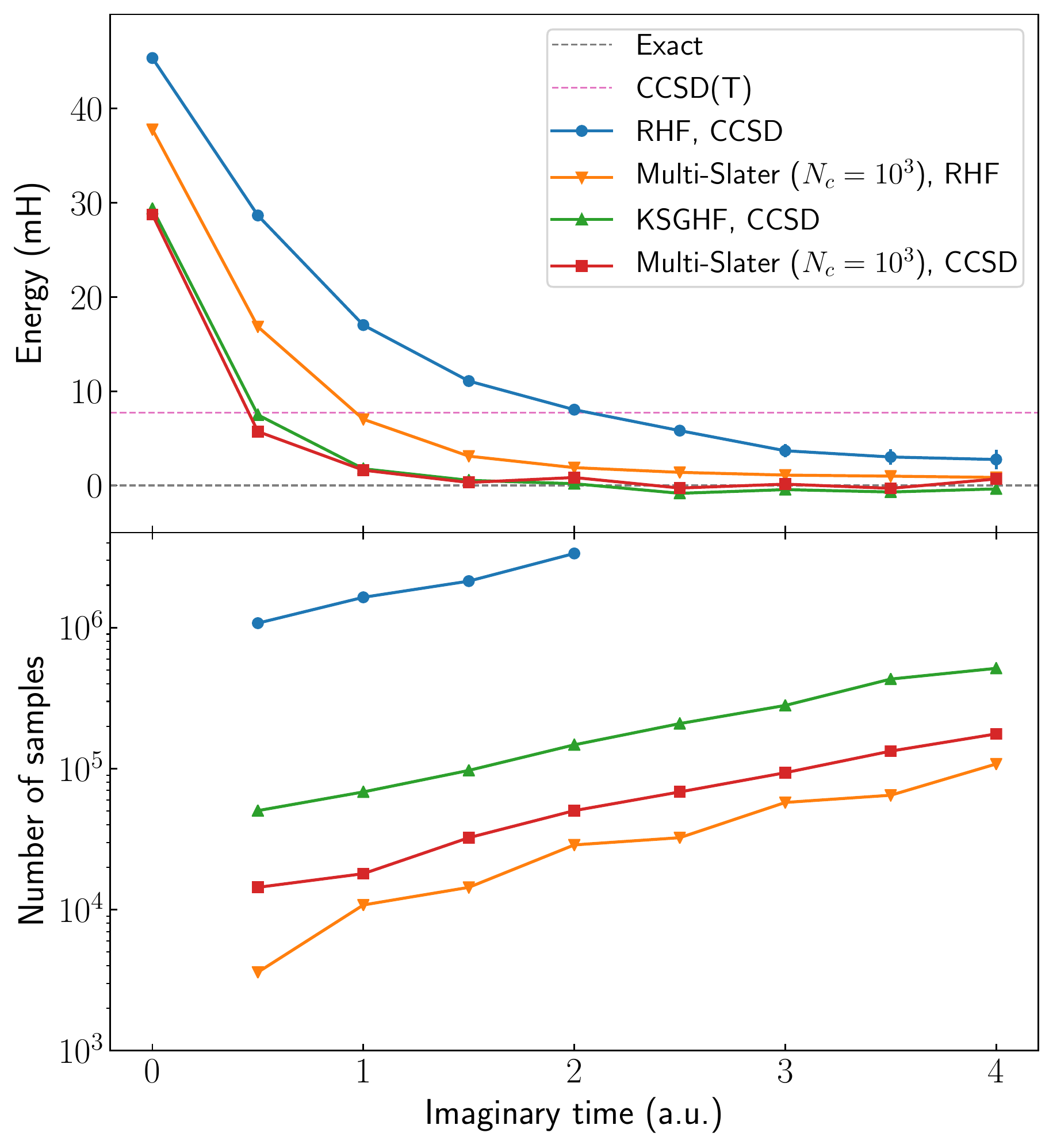}
\caption{Convergence of the energy estimator for \ce{N2} (\(d = 3\) Bohr) in cc-pVDZ basis for different trial state combinations denoted as \textit{wave}\textsubscript{1}, \textit{wave}\textsubscript{2}. For example, ``RHF, CCSD" indicates the estimator using an RHF state as \(\ket{\psi_l}\) and a CCSD state as \(\ket{\psi_r}\). The bottom plot shows the number of samples required to get a stochastic error less than \(0.5\) mH as a function of imaginary time.}\label{fig:n2}
\end{figure}

\begin{figure}[htp]
\centering
\includegraphics[width=0.45\textwidth]{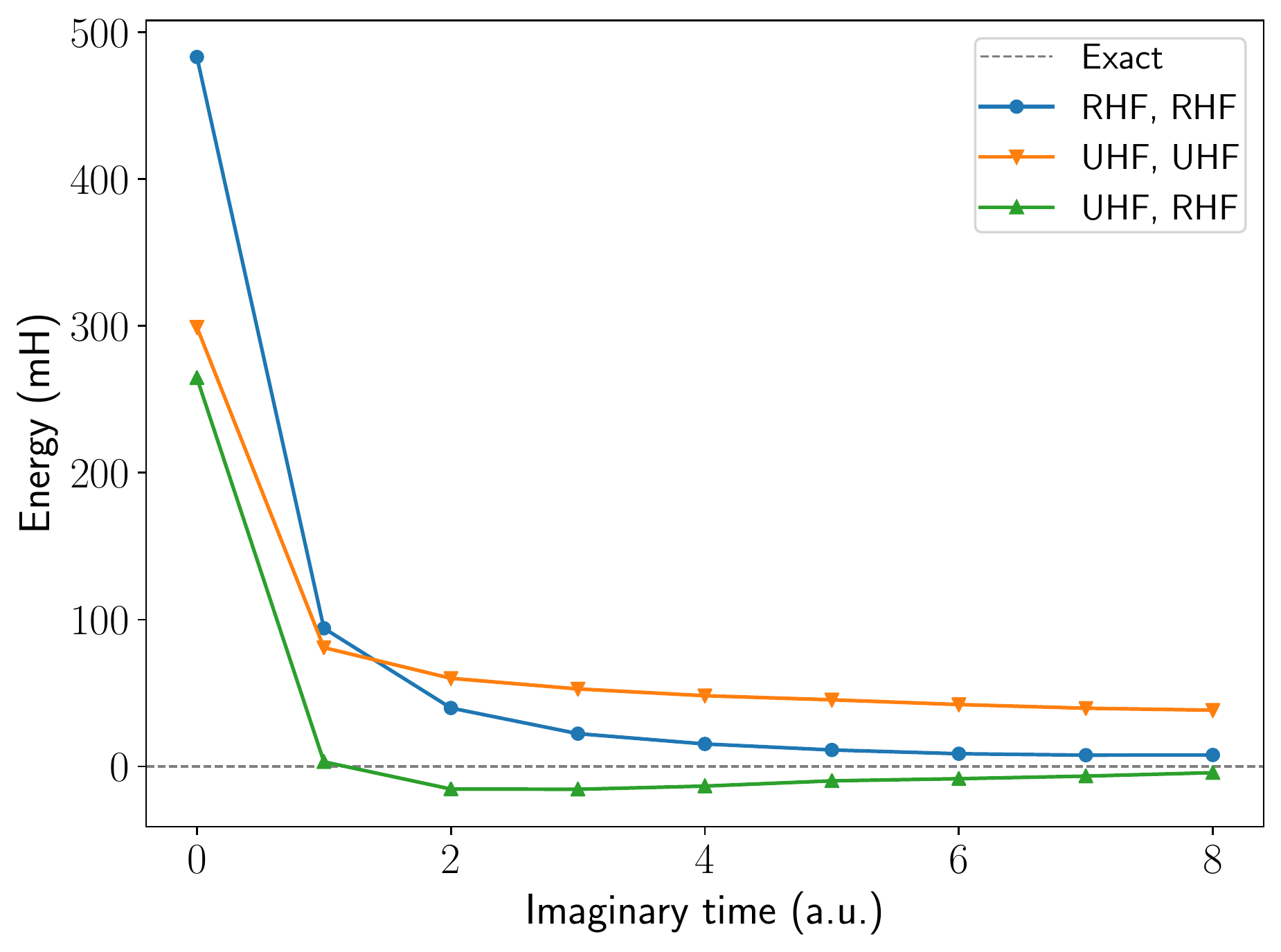}
\caption{Slow or non-variational convergence of the energy estimator for different trial state combinations. System and other details same as in Figure \ref{fig:n2}}\label{fig:n2_1}
\end{figure}
In this section, we analyze the convergence of the energy estimator in equation \ref{eq:estimator} for different combinations of the trial wave functions discussed above. To understand the convergence behavior of the function \(E(\tau)\) in equation \ref{eq:estimator} with \(\tau\), let us denote the spectral decomposition of the trial states in the basis of energy eigenstates as
\begin{equation}
   \ket{\psi_l} = \sum_i l_i \ket{\psi_i},\qquad \ket{\psi_r} = \sum_i r_i \ket{\psi_i},
\end{equation}
where \(H\ket{\psi_i} = E_i\ket{\psi_i}\). Substituting into equation \ref{eq:estimator}, we get
\begin{equation}
   E(\tau) = \frac{\sum_iE_ie^{-\tau E_i}l_i^*r_i}{\sum_ie^{-\tau E_i}l_i^*r_i}.
\end{equation}
From this expression it is evident that the convergence is dictated by the energy gap to the low-lying excited states in the symmetry sector of the ground state and the coefficients of these states in the spectral decompositions.

Figure \ref{fig:n2} shows the convergence behavior in a stretched \ce{N2} molecule with a bond length of \(d=3\) Bohr using the cc-pVDZ basis set. We correlated all 14 electrons in the 28 orbitals of this basis set for all calculations. At this geometry, the electronic structure is fairly multireference in character, as evidenced by the substantial error in the CCSD(T) energy. Despite the shortcomings of the CCSD wave function in this case, it still improves the convergence of energy significantly compared to RHF. Similarly, both multi-Slater and KSGHF wave functions used as \(\ket{\psi_l}\) trials lead to faster energy convergence than RHF. For all combinations shown in the plot, the estimators are variational throughout the propagation. Figure \ref{fig:n2_1} shows that when a UHF state is used along with an RHF state in a mixed estimator, the resulting energy becomes non-variational after some propagation and eventually converges to the ground state energy slowly. Curiously, the variational estimator using UHF for both \(\ket{\psi_l}\) and \(\ket{\psi_r}\) converges rather slowly compared to the RHF variational estimator, even though the UHF energy is lower at \(\tau=0\). Both these observations can be explained by noting that the UHF state has a smaller overlap with the ground state and excited states contribute significantly in its spectral decomposition. This example highlights the importance of employing trial states having large overlaps with the ground state, with low-lying excited states filtered out as much as possible. It also suggests active space wave functions used in quantum chemistry may serve as good trial states. In our experiments, we have found the estimator to become non-variational when using a symmetry broken wave function as one of the trial states. While such non-variational estimators could still be employed in fp-AFQMC calculations, determining convergence could become tricky for larger systems. Thus we do not use symmetry broken states like UHF or UCCSD with a UHF reference in this study.

Another important consideration is the statistical efficiency afforded by various trial states. The bottom panel of figure \ref{fig:n2} shows the number of samples required to get a stochastic error of less than 0.5 mH at different imaginary times. In all cases, the number of samples increases roughly exponentially with imaginary time because of the sign problem, but the estimators differ considerably in the number required to get a fixed stochastic error. As discussed in section \ref{sec:propagator_sampling}, based on the zero-variance principle (equation \ref{eq:zv}), statistical efficiency is dictated by the quality of the \(\ket{\psi_l}\) state. This principle is evident from the plot, with the estimator using the RHF state as \(\ket{\psi_l}\) requiring a large number of samples. The multi-Slater state, with the leading \(10^3\) configurations taken from an HCI wave function, is the most accurate and, thus, yields the most statistically efficient estimators.

\subsubsection{Multi-Slater trials with different number of configurations}\label{sec:multi_slater_convergence}
\begin{figure}[htp]
\centering
\includegraphics[width=0.45\textwidth]{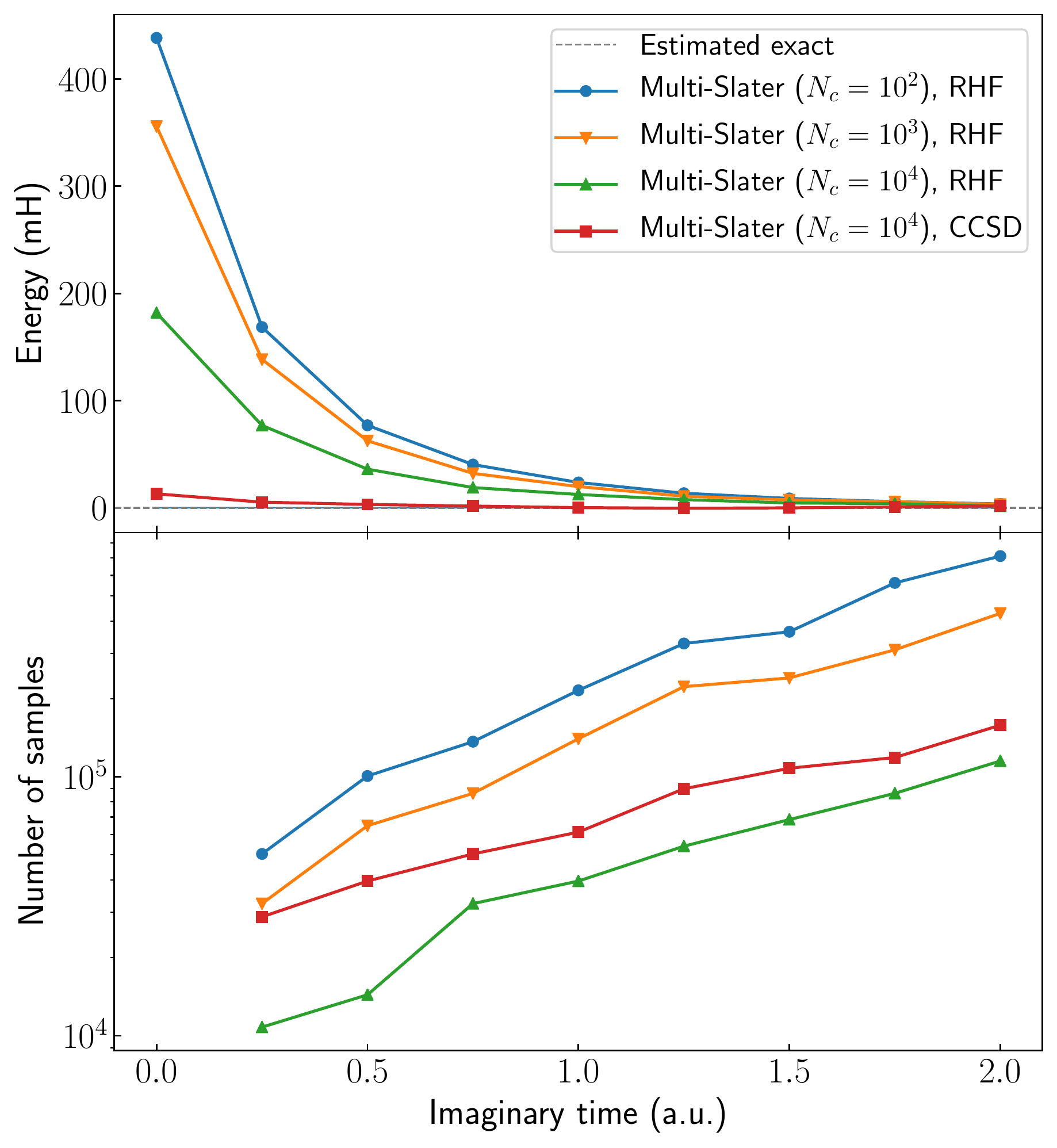}
\caption{Convergence of the energy estimator for two water molecules at equilibrium geometries in aug-cc-pVDZ basis for different number of configurations. Other details same as in Figure \ref{fig:n2}}\label{fig:h2o}
\end{figure}

\begin{figure}[htp]
\centering
\includegraphics[width=0.45\textwidth]{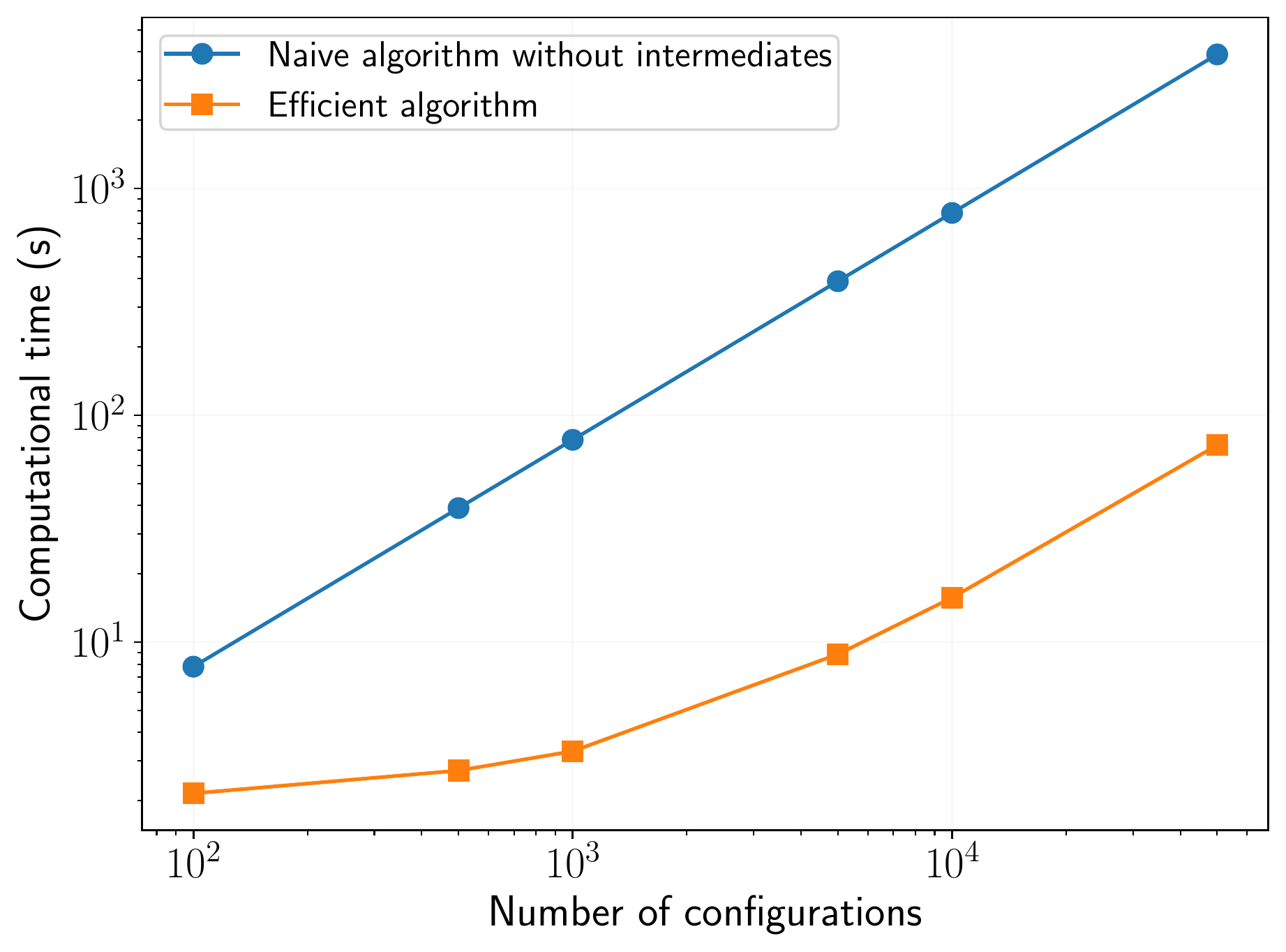}
\caption{Computational cost scaling of the local energy calculation with the number of configurations (same system as in Figure \ref{fig:h2o}). Computational time is the wall time for calculating 100 local energy samples. Efficient algorithm refers to the algorithm described in section \ref{sec:multi_slater}.}\label{fig:h2o_1}
\end{figure}

Multi-Slater trial states can be systematically made more accurate by adding more configurations. In this section, we analyze the cost and benefits of using longer expansions. Figure \ref{fig:h2o} shows energy convergence with imaginary time for two water molecules at equilibrium geometries (provided in supporting information) in the aug-cc-pVDZ basis set. The \(1s\) electrons on oxygen atoms were frozen in all calculations correlating the remaining 16 electrons in 80 orbitals. In contrast to the \ce{N2} example in the last subsection, correlation in this system is primarily dynamic. The main advantage of adding more configurations to the trial state \(\ket{\psi_l}\) is to increase the sampling efficiency, as seen from the bottom panel of the figure. It also slightly reduces the projection time required to reach convergence. We have included the curve for the estimator using a CCSD state as \(\ket{\psi_r}\) to demonstrate the efficacy of this state in accelerating convergence particularly in problems dominated by dynamic correlation.

In figure \ref{fig:h2o_1}, we compare the computational cost scaling of the naive local energy algorithm with that presented in section \ref{sec:multi_slater} for the same two water molecule system. The naive algorithm effort is estimated by simply multiplying the cost of local energy calculation for a single RHF determinant by the number of configurations. While the cost of calculating local energy for a single determinant includes the cost of evaluating the Green's function, which need not be evaluated for every configuration from scratch, it is a small fraction of the much more expensive operation of contracting the Green's function with Hamiltonian integrals. For less than \(10^3\) configurations, the cost of the efficient algorithm is seen to increase sublinearly with the number of configurations. In this regime, the cost is dominated by formation of intermediates scaling as \(O(XNM^2)\). Beyond this point, the iteration over configurations begins to dominate, scaling linearly as \(O(XN_c)\). To estimate the marginal cost of adding configurations to the trial state, one needs to compare the increase in local energy cost with reduction in the required number of samples with the associated reduction in propagation costs depending on the system.

\subsubsection{Basis set and semicore correlation focal point corrections}
\begin{table*}[htp]
\caption{Basis set corrections for ground state energies (in H) of \ce{N2} for a set of bond lengths (in Bohr). The last column lists energies obtained by adding the UCCSD(T) cc-pVQZ basis correction to the cc-pVDZ fp-AFQMC energies. Stochastic errors in the QMC calculations are less than 0.8 mH.}\label{tab:n2}
\centering
\begin{tabular}{*{12}c}
\hline
\hline
\(d\) &~~~& \multicolumn{3}{c}{cc-pVDZ} &~~& \multicolumn{5}{c}{cc-pVQZ}\\
\cline{3-5} \cline{7-11}
&& UCCSD(T) &~~& fp-AFQMC &~~& UCCSD(T) &~~& fp-AFQMC &~~& fp-AFQMC/UCCSD(T)\\
\hline
2.0 && -109.2694 && -109.2706 && -109.4592 && -109.4610 && -109.4603\\
2.4 && -109.2356 && -109.2419 && -109.4078 && -109.4142 && -109.4141\\
2.7 && -109.1506 && -109.1642 && -109.3172 && -109.3320 && -109.3307\\
3.0 && -109.0689 && -109.0897 && -109.2311 && -109.2543 && -109.2519\\
3.6 && -108.9828 && -108.9978 && -109.1372 && -109.1537 && -109.1521\\
4.2 && -108.9630 && -108.9702 && -109.1117 && -109.1192 && -109.1189\\
\hline
\end{tabular}
\end{table*}

Gaussian basis sets used in quantum chemistry are designed to provide a fast convergence for various molecular properties, including energy differences with increasing basis set size. While much of the important correlation is captured in smaller basis sets, larger basis sets are still required to obtain quantitatively accurate properties in some cases. Focal point approaches\cite{east1993heat,csaszar1998pursuit,sinnokrot2002estimates} have been developed to yield accurate results close to the continuum limit while avoiding prohibitively expensive calculations with steep-scaling methods in large basis sets. Here, we consider its use to provide basis set and semicore corrections to fp-AFQMC calculations. The basis set correction is given as
 \begin{equation}
   \begin{split}
      E(\text{fp-AFQMC}, \text{LB}) &\approx E(\text{fp-AFQMC}, \text{SB})\\
			& + E(\text{M}, \text{LB}) - E(\text{M}, \text{SB}),
   \end{split}
\end{equation}
where LB and SB refer to large and small bases, respectively and M is another method that can be used in the large basis set. We will refer to the resulting method as fp-AFQMC/M. M could be CCSD, CCSD(T), MP2, or any other technique suitable for the problem at hand. The crucial point is that it does not need to be accurate on its own; it should be good enough to capture the effects of increasing the basis size.

Table \ref{tab:n2} shows the efficacy of the basis set correction for the \ce{N2} bond breaking problem. We performed all electron fp-AFQMC ground state energy calculations of the \ce{N2} molecule for a set of bond lengths in both cc-pVDZ and cc-pVQZ basis sets. A CCSD state was used as the \(\ket{\psi_r}\) trial, while a multi-Slater wave function was used as \(\ket{\psi_l}\). Despite the poor performance of UCCSD(T) in this system, with absolute energy errors of $\sim$20 mH for some bond lengths, it provides an excellent basis set correction for fp-AFQMC.

\begin{table}[htp]
\caption{Semicore corrections for ground state energies (in H) in transition metal oxide molecules. Ne and Ar core denote energies with and without semicore correlation, respectively. All calculations were performed with the X2C Hamiltonian in a triple-\(\zeta\) quality ANO basis set. Stochastic error in QMC calculations is less than 0.7 mH in all cases.}\label{tab:semicore}
\centering
\begin{tabular}{ccccccc}
\hline
\hline
Species && Method &~& \ce{Ar} core &~& \ce{Ne} core\\
\hline
\ce{CrO} && UCCSD(T) && -1115.7928 && -1116.1485\\
&& fp-AFQMC && -1115.7953 && -1116.1505\\
&& fp-AFQMC/UCCSD(T) && - && -1116.1510 \\
\hline
\ce{MnO} && UCCSD(T) && -1232.5244 && -1232.8777\\
&& fp-AFQMC && -1232.5287 &&  -1232.8815\\
&& fp-AFQMC/UCCSD(T) && - &&  -1232.8820\\
\hline
\ce{FeO} && UCCSD(T) && -1346.7165 && -1347.0381\\
&& fp-AFQMC && -1346.7188 &&  -1347.0415\\
&& fp-AFQMC/UCCSD(T) && - && -1347.0404 \\
\hline
\end{tabular}
\end{table}

Similar to basis set corrections, a semicore correlation correction can be used as
 \begin{equation}
   \begin{split}
		 E(\text{fp-AFQMC}, N_{\text{v}}+N_{\text{sc}})&\approx E(\text{fp-AFQMC}, N_{\text{v}})\\
		 + &E(\text{M}, N_{\text{v}}+N_{\text{sc}}) - E(\text{M}, N_{\text{v}}),
   \end{split}
\end{equation}
where \(N_{\text{v}}\) is the number of valence electrons and \(N_{\text{sc}}\) is the number of semicore electrons. This correction is particularly relevant for \(3d\) transition metal compounds where in addition to valence electrons, the semicore \(3s\) and \(3p\) electrons can play an important role in the chemistry. We note that a CCSD(T) semicore correction has been used along with CASPT2 recently.\cite{phung2018toward} Table \ref{tab:semicore} shows the performance of this correction for fp-AFQMC in three transition metal oxide molecules at equilibrium geometries used in reference \onlinecite{williams2020direct}. We used the scalar relativistic X2C Hamiltonian in a triple-\(\zeta\) ANO-RCC basis set with a \(21s15p10d6f4g/6s5p3d2f1g\) contraction on the metal atoms and a \(14s9p4d3f/4s3p2d1f\) contraction on oxygen. Multi-Slater wave functions were used as \(\ket{\psi_l}\) trials. Since these systems are open shell, we used unrestricted CCSD wave functions on top of ROHF references as the \(\ket{\psi_r}\) trial states. In all three cases, UCCSD(T) energies have some absolute energy errors, but they still provide reasonable semicore corrections.

Besides the examples presented here, we will check the accuracy of these focal point corrections in larger systems in the following sections. In all the cases presented in this paper, we have found these corrections to be exceptionally accurate.

\subsection{Organic molecules}\label{sec:organic_molecules}
\subsubsection{Ground state energy of benzene (\ce{C6H6})}\label{sec:benzene}

\begin{table}[htp]
\caption{Ground state energy (in H) of benzene from different methods. Correlation spaces: (30e, 108o) for cc-pVDZ and (30e, 258o) for cc-pVTZ.}\label{tab:benzene}
\centering
\begin{tabular}{ccccc}
\hline
\hline
Method &~~~& cc-pVDZ &~~& cc-pVTZ\\
\hline
CCSD(T) && -231.5813 && -231.8058 \\
MBE-FCI\cite{eriksen2020ground} && -231.5848 && - \\
DMRG\cite{eriksen2020ground} && -231.5846(7) && - \\
SHCI\cite{eriksen2020ground} && -231.586(2) && - \\
ph-AFQMC (RHF)\cite{lee2020performance} && -231.5879(4) && -231.8122(4) \\
ph-AFQMC (multi-Slater)\cite{lee2020performance} && -231.5861(4) && - \\
fp-AFQMC/CCSD(T) && - && -231.8096(7) \\
fp-AFQMC && -231.5851(7) && -231.809(1) \\
\hline
\end{tabular}
\end{table}

A recent benchmark study\cite{eriksen2020ground} reported the ground state energy of benzene in the cc-pVDZ basis using various accurate quantum chemistry methods. Here we calculate fp-AFQMC energies in this basis as well as the cc-pVTZ basis. We used the same geometry as in the benchmark paper and froze the \(1s\) electrons of all carbon atoms at the HF level. This yields correlation spaces of (30e, 108o) for the cc-pVDZ basis and (30e, 258o) for the cc-pVTZ basis. For fp-AFQMC calculations in both basis sets, we used a multi-Slater wave function as \(\ket{\psi_l}\) and the CCSD wave function as \(\ket{\psi_r}\). In the cc-pVDZ basis calculation, we used the canonical HF orbitals to obtain an HCI wave function  consisting of about \(2.6 \times 10^5\) configurations, of which the leading \(10^4\) were used in the trial state. For the cc-pVTZ basis set, we first performed an HCISCF calculation with a (30e, 100o) active space with a loose \(\epsilon_1=0.001\) allowing internal rotations. An HCI wave function consisting of about \(1.3\times 10^6\) configurations was then obtained in just the active space, and the leading \(10^4\) configurations were used in the trial state for fp-AFQMC. Table \ref{tab:benzene} shows the ground state energies from various methods. The correlation in this system is primarily dynamic in nature and, thus, CCSD(T) recovers most of the correlation energy. All the high accuracy methods are more or less in agreement within chemical accuracy for the cc-pVDZ basis. In the cc-pVTZ basis, the phaseless approximation error with the RHF trial is about 3 mH. The CCSD(T) basis set correction seems to work well in this case as well. We note that unlike several methods in table~\ref{tab:benzene}, the fp-AFQMC energies were obtained without performing any extrapolations.

\subsubsection{Automerization of cyclobutadiene (\ce{C4H4})}\label{sec:cyclobutadiene}

\begin{figure}[htp]
\centering
\includegraphics[width=0.4\textwidth]{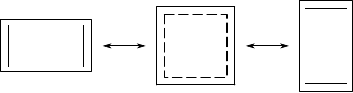}
\caption{\(D_{4h}\) symmetric transition state of cyclobutadiene (middle) between two \(D_{2h}\) symmetric degenerate minima (left and right)}\label{fig:cbd}
\end{figure}

Calculating the barrier height for automerization of cyclobutadiene is a challenging problem because of the multireference (biradical) character of the transition state. This system has been used to gauge the accuracy of many multireference methods. We performed fp-AFQMC calculations using geometries provided in reference \onlinecite{lyakh2011tailored}. The minima correspond to a \(D_{2h}\) rectangular geometry, while the transition state has a square \(D_{4h}\) geometry (figure \ref{fig:cbd}). Some geometry optimization studies find slightly bent transition state structures,\cite{dang2021fully} but due to the flatness of the energy surface near the transition state, energy differences between different geometries used for transition states in various studies are tiny (less than a mH in most cases). Carbon \(1s\) orbitals were frozen, leading to correlation spaces of size (20e, 72o) and (20e, 172o) in the cc-pVDZ and cc-pVTZ basis sets, respectively. We used the CCSD wave function as the \(\ket{\psi_r}\) trial state in both bases. For \(\ket{\psi_l}\), we used multi-Slater wave functions consisting of the order of \(10^4\) configurations were obtained from relatively crude HCI calculations using a procedure similar to that used for benzene calculations. Full details of the wave functions can be found in the SI.

\begin{table}[htp]
\caption{Automerization barrier height (kcal/mol) for cyclobutadiene. Note that different methods used slightly different geometries. Correlation spaces: (20e, 72o) for cc-pVDZ and (20e, 172o) for cc-pVTZ.}\label{tab:cyclobutadiene}
\centering
\begin{tabular}{ccccc}
\hline
\hline
Method &~~~& cc-pVDZ &~~& cc-pVTZ\\
\hline
CCSD(T) && 15.8 && 18.2 \\
CC(t;3)\cite{shen2012combining} && 7.8 && 10.0 \\
CCSDT\cite{shen2012combining} && 7.6 && 10.6 \\
TCCSD (2, 2)\cite{lyakh2011tailored} && 9.4 && 12.9\\
TCCSD(T) (2, 2)\cite{lyakh2011tailored} && 4.6 && 7.0 \\
TCCSD (12, 12)\cite{vitale2020fciqmc} && - && 9.2 \\
MR-MkCCSD(T)\cite{bhaskaran2008multireference} && 7.8 && 8.9 \\
MRCI+Q\cite{vitale2020fciqmc} && - && 9.2 \\
fp-AFQMC/CCSD(T) && - && 10.9(4) \\
fp-AFQMC && 8.4(4) && 10.2(4) \\
\hline
iCAS-CI (6-31+G** basis)\cite{dang2021fully} && 11 && \\
\hline
Experiment\cite{whitman1982limits} && 1.6-10 && \\
\hline
\end{tabular}
\end{table}

Table \ref{tab:cyclobutadiene} shows barrier heights obtained from different methods. We also provide absolute energies in the SI for reference. The experimental estimate for the barrier spans a large range of 1.6-10 kcal/mol (includes zero-point vibrational energy). CCSD(T) overestimates the gap substantially due to its poor description of the transition state. An iterative treatment of triples excitations in CCSDT or its cheaper approximation CC(t;3)\cite{shen2012combining} almost entirely corrects this error, signaling the failure of perturbative treatment of triples to describe the biradical transition state. Tailored CCSD (TCCSD)\cite{lyakh2011tailored} with a small (2e, 2o) active space reference overestimates the barrier height, while the perturbative triples correction overcorrects this error. Recent TCCSD calculations with a (12e, 12o) CAS reference\cite{vitale2020fciqmc} improve upon the previous result. The genuinely multireference Mukherjee's CCSD (MR-MkCCSD(T))\cite{bhaskaran2008multireference} yields a barrier height within about 1 kcal/mol of the fp-AFQMC result. Multireference configuration interaction with the Q correction (MRCI+Q) and incremental complete active space configuration interaction (iCAS-CI)\cite{dang2021fully} also give energy gaps close to the fp-AFQMC value. Finally, despite the poor performance of CCSD(T), it provides a remarkably accurate basis set correction for fp-AFQMC.

\subsection{Isomerization of \ce{Cu2O2^2+}}\label{sec:cu2o2}
\begin{figure}[htp]
\centering
\includegraphics[width=0.45\textwidth]{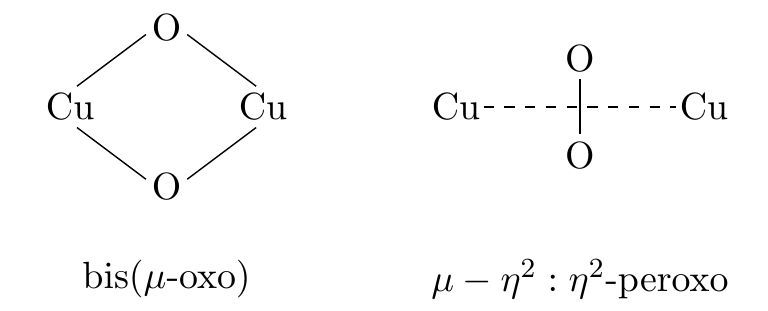}
\caption{\ce{Cu2O2^2+} isomers corresponding to \(f=0\) (left) and \(f=1\) (right)}\label{fig:cu2o2_struct}
\end{figure}

Enzymes like tyrosinase and catechol oxidase feature a binuclear \ce{Cu2O2^2+} active site. The copper atoms serve to activate molecular oxygen and can bind to it in a variety of ways.\cite{solomon1992electronic} A particular reaction pathway connecting two such structural isomers shown in figure \ref{fig:cu2o2_struct} has been studied extensively using several wave function methods and DFT functionals.\cite{cramer2006theoretical,malmqvist2008restricted,marti2008density,kurashige2009high,yanai2010multireference,liakos2011interplay,jimenez2013multi,landinez2019non,blunt2020efficient} It is known to be a challenging electronic structure problem with a large spread in calculated isomerization energetics obtained from different techniques. While a thorough study of this system would necessitate the inclusion of solvation and ligand effects, in this paper, we only focus on the core and compare various methods with the results obtained using fp-AFQMC. We also gauge the impact of semicore correlation and scalar relativity on relative energies.

\begin{figure}[htp]
\centering
\includegraphics[width=0.45\textwidth]{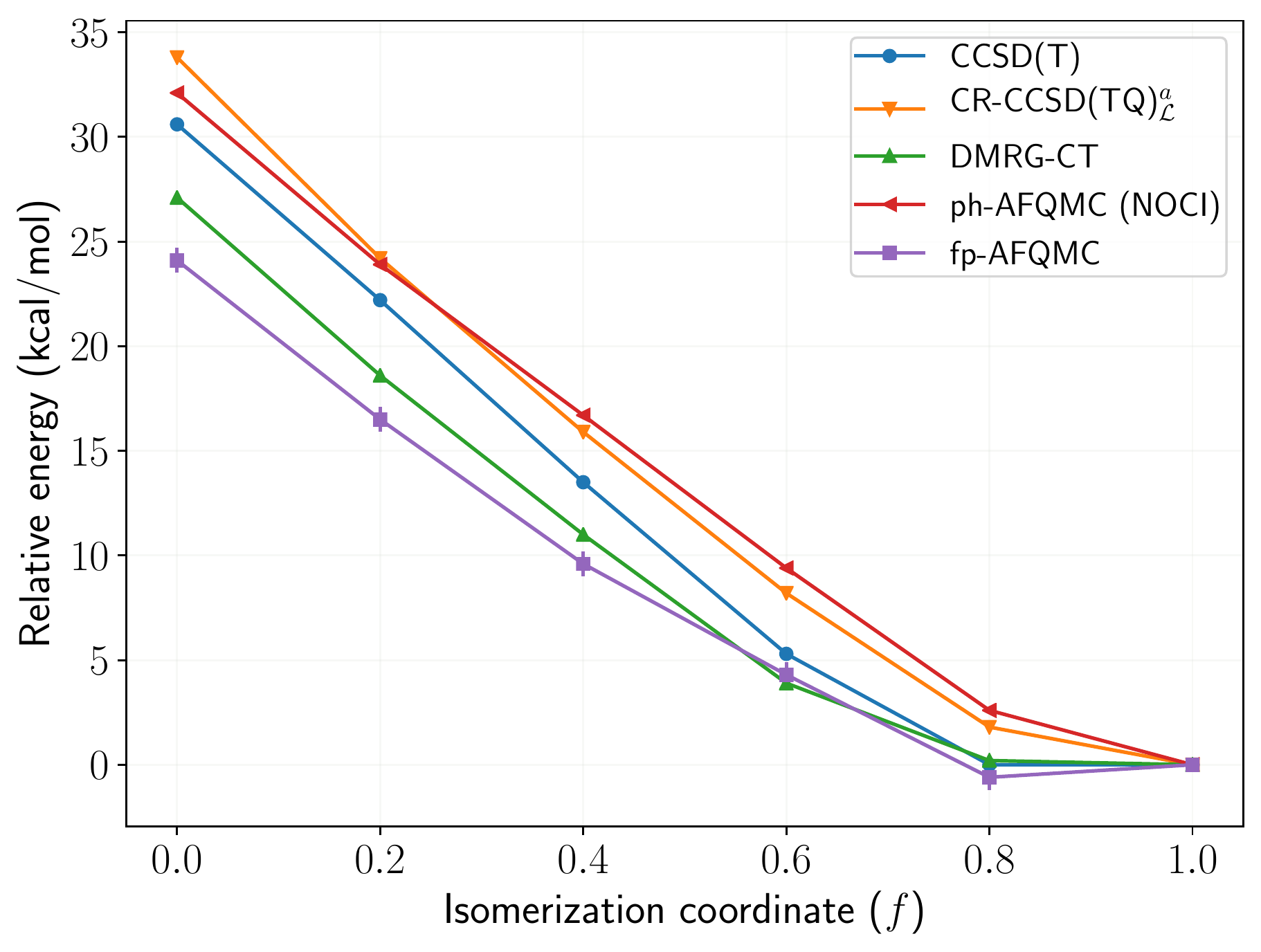}
\caption{Relative energies along the isomerization curve of \ce{Cu2O2^2+}.}\label{fig:cu2o2_curve}
\end{figure}

\begin{table*}[htp]
\caption{Relative energy \(\Delta E = E(f=0) - E(f=1)\) (in kcal/mol) for \ce{Cu2O2^2+}. Refer to the main text for full descriptions of the basis sets.}\label{tab:cu2o2}
\centering
\begin{threeparttable}
\begin{tabular}{ccccccccc}
\hline
\hline
Basis set type &~~& Relativistic effects &~~& Method &~~& Correlation space &~~& \(\Delta E\)\\
\hline
\ce{Cu}: Stuttgart ECP && Included in && CCSD(T) && (32e, 108o) && 30.6 \\
\ce{O}: ANO triple-\(\zeta\) && the ECP && CR-CCSD(TQ)\(^a_{\mathcal{L}}\)\cite{cramer2006theoretical} && (32e, 108o) && 33.8 \\
(BS1) && && DMRG-CT\cite{yanai2010multireference} && (32e, 108o) &&  27.1\\
&& && ph-AFQMC (NOCI)\cite{landinez2019non} && (32e, 108o) && 32.1 \\
&& && fp-AFQMC && (32e, 108o) &&  24.1(6) \\
\\
&&  && CCSD(T)  && (52e, 118o) &&  34.5\\
&& && \(D_{2h}\)S-UHF [8 det]\cite{jimenez2013multi} && (52e, 118o) && 50\tnote{*}\\
&& && ph-AFQMC (NOCI)\cite{landinez2019non} && (52e, 118o) && 41.9 \\
&& && fp-AFQMC && (52e, 118o) &&  31.0(6) \\
\hline
ANO triple-\(\zeta\) && X2C && CCSD(T) && (32e, 158o) && 34.8 \\
(BS2) && && CCSD(T) && (52e, 166o) && 33.6\\
&& && fp-AFQMC && (32e, 158o) && 29(1)\\
\hline
ANO quadruple-\(\zeta\) && X2C && CCSD(T) && (52e, 290o) && 33.0\\
(BS3) && && fp-AFQMC/CCSD(T) && (52e, 290o) && 27(1)\\
\hline
ANO quadruple-\(\zeta\) && Not included && CCSD(T) && (32e, 280o) && 37.3\\
(BS3-NR) && && CCSD(T) && (52e, 290o) && 38.1\\
&& && SC-NEVPT2\cite{blunt2020efficient} && (48e, 288o) && 41.3(8)\\
\hline
\end{tabular}
\begin{tablenotes}
	\item[*] estimated from the plot in reference \onlinecite{jimenez2013multi}.
\end{tablenotes}
\end{threeparttable}
\end{table*}

The isomerization pathway is obtained by simply linearly interpolating between the two structures shown in figure \ref{fig:cu2o2_struct} using a single parameter \(f\) as described in reference \onlinecite{cramer2006theoretical}. To allow a direct comparison with previous studies, we performed fp-AFQMC calculations using the same double-\(\zeta\) quality basis that has often been employed, referred to here as BS1. This basis used the Stuttgart pseudopotential and associated basis functions for the copper atoms, while an ANO basis with \(10s6p3d/4s3p2d\) contraction for the oxygen atoms. We froze the semicore \(3s\) and \(3p\) electrons on copper and the core \(1s\) electrons on oxygen at the HF level leading to a correlation space (32e, 108o). Multi-Slater and CCSD wave functions were used as \(\ket{\psi_l}\) and \(\ket{\psi_r}\) trial states, respectively (details in the SI). The calculated relative energies from fp-AFQMC and a few other methods are shown in figure \ref{fig:cu2o2_curve}. Note that all the methods shown used the same basis set. The correlation space was the same for all techniques except DMRG-CT (strongly contracted canonical transformation), which froze the same number of orbitals as other methods but at the DMRG-SCF level. We also list the relative energies between the two isomers in table \ref{tab:cu2o2}. CCSD(T) overstabilizes the \(\mu-\eta^2:\eta^2\)-peroxo isomer by about 6.5(6) kcal/mol compared to fp-AFQMC. Looking at the absolute energies (provided in SI), this difference is mainly due to CCSD(T) undercorrelating the bis(\(\mu\)-oxo) isomer by about 8 mH. On the other hand, the perturbative treatment of triples in CCSD(T) seems to provide a good description of the correlation in the \(\mu-\eta^2:\eta^2\)-peroxo isomer, slightly overcorrelating it, suggesting that this state does not have a significant biradical character. CR-CCSD(TQ)\(^a_{\mathcal{L}}\) and ph-AFQMC with a nonorthogonal CI trial both overstabilize the \(\mu-\eta^2:\eta^2\)-peroxo isomer even further. The DMRG-CT curve is in relatively good agreement with the fp-AFQMC results.

To understand the effect of semicore correlation and the basis set on the relative energies of bis(\(\mu\)-oxo) and \(\mu-\eta^2:\eta^2\)-peroxo isomers, we performed additional calculations reported in table \ref{tab:cu2o2}. Correlating the copper \(3s\) and \(3p\) semicore electrons and the oxygen \(1s\) electrons, leading to a correlation space of (52e, 118o), substantially increases the gap in BS1. This is also true of CCSD(T) energies. We also performed fp-AFQMC calculations in the bigger triple-\(\zeta\) ANO-RCC basis set with a \(21s15p10d6f4g/6s5p3d2f1g\) contraction on copper and \(14s9p4d3f/4s3p2d1f\) on oxygen using the X2C scalar relativistic Hamiltonian. We will refer to this basis set as BS2. The fp-AFQMC gap without semicore correlation in BS2 is larger than the BS1 gap. CCSD(T) calculations suggest a much smaller impact of semicore correlations on the isomerization barrier in BS2 compared to BS1. This could be due to an inadequate representation of semicore electrons and scalar relativistic effects in BS1. Previous studies have noted the sizeable influence of scalar relativity on relative energies as well as their poor description in the ECP of BS1.\cite{kurashige2009high,liakos2011interplay} CCSD(T) calculations in the quadruple-\(\zeta\) quality ANO-RCC basis set (BS3) suggest the relative energy is nearly converged with the basis set size for BS3. The fp-AFQMC/CCSD(T) energy was obtained by adding semicore and basis set corrections to the fp-AFQMC BS3 value and is the best estimate of the relative energy presented here. Non-relativistic CCSD(T) and strongly contracted n-electron valence perturbation theory (SC-NEVPT2) calculations performed in the quadruple-\(\zeta\) ANO basis set (termed BS3-NR), also reported in table \ref{sec:cu2o2}, confirm the importance of including scalar relativity in this system. Relative energies obtained from non-relativistic calculations are significantly larger than relativistic ones, while semicore correlation effects are relatively small for CCSD(T). Based on this evidence, it seems likely that semicore correlation effects are relatively small in this system provided an appropriate basis set is employed.

\section{Conclusion}
We have presented efficient ways to use high-quality trial wave functions, namely multi-Slater, CCSD, and symmetry projected mean-field states, in near-exact free projection AFQMC. Appropriate combinations of trial states reduce the amount of projection time required to achieve convergence and mitigate the sign problem. We analyzed how the trial states affect the statistical efficiency of Monte Carlo sampling using illustrative examples. We also found some encouraging evidence suggesting focal point basis set and semicore corrections may allow widening the scope of this approach when used in combination with other methods. We provided benchmark energy values for the challenging problems of automerization of cyclobutadiene and isomerization of \ce{Cu2O2^2+} in moderately sized basis sets.

The method presented here can be improved in several ways that we plan to pursue in the future. To tackle systems with low lying excited states that are difficult to project out deterministically using the wave functions presented here, a transcorrelated method with trial states including Jastrow factors may be effective. Another possibility is searching for constraints like the phaseless approximation that are better controlled. Our experiments with the systems considered here suggest that direct Gaussian sampling is very efficient for moderately correlated systems. Importance sampling is likely necessary for more strongly correlated systems. A more systematic study of focal point corrections is warranted to assess their scope of applicability. It will also be fruitful to explore alternative ways to extrapolate energies to the basis set limit.

Improvements to the software implementation along the lines of those already introduced for ph-AFQMC are possible. Significant cost reductions can be gained by making full use of symmetries in the \textit{ab initio} Hamiltonian.\cite{motta2019hamiltonian} Our current implementation only takes advantage of the spin symmetry. The use of graphical processing units has been suggested to provide drastic speedups in AFQMC calculations.\cite{shee2018phaseless,malone2020accelerating} Such performance improvements should allow applications to larger systems than those presented here at a fraction of the cost. In addition to performance improvements, we are also interested in looking for ways to calculate properties besides ground state energies like correlation functions and nuclear forces within the framework presented here.

\section*{Acknowledgements}
The funding for this project was provided by the national science foundation through the grant CHE-1800584. SS was also partly supported through the Sloan research fellowship. All calculations were performed on the Blanca and Summit clusters at CU Boulder. We thank Takeshi Yanai for providing the basis set information and Miguel Morales for the ph-AFQMC energies for \ce{Cu2O2^2+}. We thank Joonho Lee for useful discussions.

\providecommand{\latin}[1]{#1}
\makeatletter
\providecommand{\doi}
  {\begingroup\let\do\@makeother\dospecials
  \catcode`\{=1 \catcode`\}=2 \doi@aux}
\providecommand{\doi@aux}[1]{\endgroup\texttt{#1}}
\makeatother
\providecommand*\mcitethebibliography{\thebibliography}
\csname @ifundefined\endcsname{endmcitethebibliography}
  {\let\endmcitethebibliography\endthebibliography}{}

\end{document}